\documentclass{pasj00}
\draft

\begin{document}
\SetRunningHead{Kaneda et al.}{AKARI telescope}
\Received{//}
\Accepted{//}

\title{In-orbit focal adjustment of the AKARI telescope with IRC images}

\author{%
   Hidehiro \textsc{Kaneda},\altaffilmark{1}
   Woojung \textsc{Kim},\altaffilmark{1}
   Takashi \textsc{Onaka},\altaffilmark{2}\\
   Takehiko \textsc{Wada},\altaffilmark{1}
   Yoshifusa \textsc{Ita},\altaffilmark{1}
   Itsuki \textsc{Sakon},\altaffilmark{2}
   and
   Toshinobu \textsc{Takagi},\altaffilmark{1}}
 \altaffiltext{1}{Institute of Space and Astronautical Science, \\
Japan Aerospace Exploration Agency, Sagamihara, Kanagawa 229-8510}
 \email{kaneda@ir.isas.jaxa.jp}
 \altaffiltext{2}{Department of Astronomy, Graduate School of Science, University of Tokyo, \\
Bunkyo-ku, Tokyo 113-0003}



\maketitle

\begin{abstract}
AKARI currently in space carries onboard a cryogenically-cooled lightweight telescope with silicon carbide mirrors. The wavefront error of the AKARI telescope obtained in laboratory measurements at 9 K showed that expected in-orbit imaging performance was diffraction-limited at a wavelength of 6.2 $\mu$m. The AKARI telescope has a function of focus adjustment by shifting the secondary mirror in parallel to the optical axis. On the 4th day after the jettison of the cryostat aperture lid in the orbit, we observed a star with the InfraRed Camera (IRC) onboard AKARI. Since the initial star images observed in the near-infrared (IR) bands were significantly blurred, we twice moved the secondary mirror for the focal adjustment based on the results of model analyses as well as data analyses of the near-IR images. In consequence, we have successfully adjusted the focus of the telescope. The in-orbit imaging performance thus obtained for the AKARI telescope is diffraction-limited at a wavelength of 7.3 $\mu$m, slightly degraded from that expected from the laboratory measurement.    
\end{abstract}

\section{Introduction}
AKARI (formerly called ASTRO-F; Murakami et al. 2007) currently in space carries onboard a cryogenically-cooled lightweight telescope, which forms an F/6 Ritchey-Chretien system with the primary mirror of 685 mm in effective diameter. The telescope is specified to be diffraction-limited at a wavelength of 5 $\mu$m at the operating temperature of 6 K. The primary and seconday mirrors are made of silicon carbide (SiC). These are the first SiC mirrors that have ever been in use for space cryogenic application in the world. Following the AKARI SiC telescope, the Herschel Space observatory with the 3.5-m-diameter SiC telescope is planned to be launched in 2008 (Pilbratt 2004; Toulemont et al. 2004). Among various kinds of SiC material, we have adopted sandwich-type SiC that consists of a porous core and a chemical-vapor-deposited coat of SiC on the surface. Details of the telescope design are described in Onaka et al. (1998), and details of the mirror fabrication and telescope integration processes are reported in Kaneda et al. (2003b). Results of cryogenic optical testing of the sandwich-type SiC mirrors and the AKARI flight telescope are described in Kaneda et al. (2003a) and Kaneda et al. (2005), respectively. The wavefront error of the AKARI telescope system obtained by the interferometric measurement at 9 K showed that expected in-orbit imaging performance was diffraction-limited at a wavelength of 6.2 $\mu$m, which was a little worse than our original goal of diffraction-limited performance at 5.0 $\mu$m. 

The AKARI telescope has a function of focal adjustment; the secondary mirror can be shifted in parallel to the optical axis by a cryogenic stepping motor through leaf springs, so that the distance between the primary and the secondary mirror is adjustable within a stroke of 0.76 mm and with a resolution of 0.466 $\mu$m, which are equivalent to 21.6 mm and 0.0131 mm at the telescope focal plane. The characteristics of the secondary focus stepping motor are listed in table 1. We operated the stepping motor with as small phase current and as fast pulse speed as possible to suppress thermal influx to the telescope system.  Prior to the launch, we conducted focal adjustment of the telescope at 9 K during laboratory measurements using a liquid-helium test chamber, and thus we anticipated that the telescope would be in best focus when it cooled to the operating temperature in the orbit. However, since the ground-based adjustment had larger uncertainties than the required in-orbit focus accuracy due to differences between the test chamber and in-orbit optical configuration, we had to be prepared for refocusing the telescope in the orbit. 

The Spitzer Space Telescope succeeded in its refocusing in 2003 after arrival of the spacecraft in orbit (Gehrz et al. 2004). Nevertheless, it is still quite challenging to safely perform an accurate focus adjustment of a cryogenic telescope in space. Prior to the launch, we had checked the normal move of the secondary mirror within a full or partial stroke at $4-6$ K in the 8 seperate cooling runs of the satellite system test. Because of concern about a failure in the cryogenic adjustment mechanism affecting all the AKARI mission, however, we did not intend to perform a conventional focus sweep to determine the best position of the secondary mirror. In addition, we required to make the number of focus settings as small as possible and to finish the adjustment as soon as possible due to a limited amount of liquid helium ($\sim 170$ liter) loaded into the AKARI cryostat (Nakagawa et al. 2007), which is only about one third of that for the Spitzer Space Telescope. 

On the 4th day after the jettison of the cryostat aperture lid when the temperature of the telescope and the focal plane instruments became stable at 6 K, we observed a star with the InfraRed Camera (IRC; Onaka et al. 2007) that is one of the focal plane instruments onboard AKARI. On the basis of blurred near-InfraRed (IR) images of the star, we decided to move the secondary mirror. The desired direction of the move and the focus setting were determined by using an optical model of the IRC and telescope together with data analyses of observed images as explained below. 

\section{Observations}
On April 17th during the AKARI performance verification phase, we observed a field star in the South Ecliptic Pole region to measure an in-orbit Point Spread Function (PSF) with the near-IR channels of the IRC, which consist of the following photometric bands: N2 covering a wavelength range of $1.7-2.7$ $\mu$m, N3 of $2.7-3.7$ $\mu$m, and N4 of $3.7-5.5$ $\mu$m. A selected star is imaged in the center of the field-of-view of each band; the image obtained in the N2 band with exposure time of 45 seconds is shown in figure 1a. The image size is $32\times 32$ pixels or $47''\times 47''$. As seen in the figure, the in-orbit initial image clearly exhibited a defocus ring and thus the telescope was found to be significantly out of focus. Therefore we definitely required to refocus the telescope in the orbit; following the above strategy, we planned to move the secondary mirror twice in two days to bring the telescope into focus. Hence, on April 18th, we made the first exploratory move with 50 secondary focus motor steps and checked the image data to find that the secondary mirror moved to the desired direction. On April 19th, we made the final move with 80 steps and again checked the resultant images to find that the focus of N2 was set to be within an acceptable range of the best focus. For each measurement of the star image, we waited for more than 4 hours after moving the secondary mirror so that the telescope system was thermally (and thus optically) stabilized on the basis of the results of laboratory measurements. The star image obtained after the final refocusing for the N2 band is shown in figure 2a.

\section{Results}
\subsection{Focus Determination}
There was ambiguity about which side of the best focus the telescope was on, when we obtained the initial blurred images of the star. In the laboratory measurement of the IRC subassembly at the operating temperature, we have found that there are residual color aberrations among the near-IR bands due to small differences between the designed refraction indices of lenses and those measured at cryogenic temperatures (Kim et al. 2003). By making use of this undesired aberrations, however, we can derive the direction of the telescope focus from the star images in the 3 near-IR bands; we have evaluated the image sizes using the following equation: 
\begin{equation}
r_{\rm mean} = \sum_{i=1}^{n}(f_i\times r_i) / \sum_{i=1}^{n}f_i,
\end{equation}
where $f_i$ is the flux falling on the $i$-th pixel and $r_i$ is the distance from the center of the image to the $i$-th pixel. The calculation was performed for $32\times 32$ pixels around the center of the image. The resultant $r_{\rm mean}$ for each near-IR band is listed in table 2, where we can see significant differences in the image size among the near-IR bands. The focal offsets of the N3 and N4 bands with regard to the focus of the N2 band, which have been obtained in the above laboratory measurement, are also listed in the table. On the basis of the order of the image size among the 3 near-IR bands, we have confirmed that the focus of the telescope is located on the far side of the designed position from the primary mirror, which implies that we have to increase the distance between the primary and the secondary mirror for the focus adjustment. The degree of defocus in the N2 band relative to those in the N3 and N4 bands is also consistent with the above focus offsets measured in the laboratory.

As a next step, we determined the optimum number of pulse steps for the move of the secondary mirror for the focal adjustment by comparing model images to measured images as performed in Hoffman et al. (2004) for the Spitzer Space Telescope.
We created both the AKARI telescope and IRC optical models in the ZEMAX ray-trace program (Kim et al. 2005). In the optical model, we have assumed that the total wavefront error of the telescope measured on the ground (Kaneda et al. 2005) is attributed to the surface deformation of the primary mirror surface, while the secondary mirror and the IRC optics have designed surface figures. Figure 1b shows the model image generated so that the image is out of focus similarly to the measured image (Fig.1a); three bright spots in the defocus ring are due to the trefoil deformation of the primary mirror observed at cryogenic temperatures (Kaneda et al. 2005). The measured image, however, appeared to suffer a coma aberration much more than the model image, and thus we managed to reproduce the measured image by tilting the primary mirror by $50''$ in the optical model (Fig.1c), which resulted in a fair agreement in shape and structure between the model and measured images. Such model images were then prepared for a variety of focal positions and cross-correlation between the model and measured images were evaluated by using the following equation: 
\begin{equation}
Crosscor = \sum_{i=1}^{n}(f_i\times simf_i)/\sqrt{\sum_{i=1}^{n}f_i^2\times\sum_{i=1}^{n}simf_i^2},
\end{equation}
where $simf_i$ is the $i$-th-pixel signal of a simulated image. This calculation results in a single number for the quality of the fit that is 1 for a perfect fit and 0 for no overlap of the images. Figure 3a shows the result of the calculation for the star image initially obtained in the AKARI orbit, which indicates that we need $\sim 130$ pulse steps for the desired focus setting of the secondary mirror; in the plot, there is a peak on either side of the telescope best focus, however, we know which side of the best focus the telescope is on from the above analysis. As a first step, we shifted the secondary mirror by 50 steps in the above direction and then performed the same calculation of the cross-correlation for the obtained image. The result is shown in figure 3b, from which we confirm that the secondary mirror has moved in the desired direction by expected amounts. Hence we finally moved the secondary mirror by another 80 steps to obtain the star image in the N2 band as shown in figure 2a; as a result, the focus of the N2 band was expected to be a little over-corrected by about 0.1 mm from the designed focal position. This focus setting has been selected to optimize the overall imaging performance of the 3 near-IR channels. Changes of relative focal positions of the 3 near-IR bands along with the focal adjustment operation are schematically shown in figure 4. 

We have shifted the secondary mirror by 61 $\mu$m in total, which is equivalent to 1.7 mm at the telescope focal plane. From figure 3a, the initial focal position of the telescope is estimated to be as much as 1.6 mm off the designed position along the optical axis in the orbit. The cause of such defocus as well as the coma aberration observed in the initial star image is yet to be investigated. Since we have conducted the focal adjustment of the telescope prior to the launch, the optical misalignment in the orbit is rather worse than expected. The most probable cause of the defocus is the cryogenic deformation of the auto-collimation flat mirror in the test chamber; we have evaluated the cryogenic deformation of the flat mirror alone by independently shifting the mirror in the chamber (Kaneda et al. 2005; Sugiyama et al. 2002). However, uncertainties in the position of the telescope focus due to non-negligible curvature of the flat mirror are on the order of 1 mm at cryogenic temperatures. The latter misalignment does not necessarily imply that the primary or the secondary mirror is really tilted; it could arise from uncertainties in the mechanical alignment of the IRC to the focal base-plate.
   
\subsection{In-orbit imaging performance}
Now that the N2 band is near best focus, the star image in the N4 band exhibits significant elongation possibly generated from the astigmatism of the telescope with the focus offset between the N2 and N4 bands, as shown in figure 2b. Therefore, we have evaluated the imaging performance of the AKARI telescope by the PSF in the N2 band. Figure 5 shows the encircled energy of the N2 PSF measured after the final focus adjustment; the half-power radius of the encircled energy is 61 $\mu$m (1 pixel = 30 $\mu$m), while the value estimated from the cryogenic optical testing of the telescope alone on the ground is 53.1 $\mu$m at a wavelength of 5 $\mu$m. By considering the 1 $\sigma$ pointing stability of the telescope that is estimated to be about 0.7 pixels in the N2 band from independent observations, the above half-power radius can be reduced down to 57 $\mu$m, which is still larger than the value on the ground. Hence the quality of the PSF is a little degraded from that expected from the laboratory measurement.

On the other hand, according to the above optical model constructed from the N2 data measured in the orbit, the wavefront error in best focus is estimated to be 0.119 $\lambda$ rms ($\lambda$: 5 $\mu$m) for the telescope alone and 0.130 $\lambda$ rms for the telescope combined with the IRC optics at the center of each field-of-view, both at a wavelength of 5 $\mu$m. As a result, the Strehl ratio of the PSF at 5 $\mu$m is 0.57 for the telescope alone and 0.51 for the telescope with the IRC. On the other hand, the wavefront error and the Strehl ratio expected from the laboratory measurement are 0.096 $\lambda$ rms and 0.69,respectively. Hence we can conclude that the in-orbit reality of the imaging performance of the telescope is diffraction-limited at a wavelength of 7.3 $\mu$m on the basis of the estimated Strehl ratio (i.e., at which wavelength the Strehl ratio becomes 0.8), a little worse than 6.2-$\mu$m-diffraction-limited imaging performance expected from the laboratory measurement, which is consistent with the above degraded PSF measured in the orbit. 

\section{Concluding Remarks}
We have successfully adjusted the focus of the AKARI telescope in the orbit by shifting the secondary mirror along the optical axis; since the initial star images observed in the near-IR bands of the IRC were significantly blurred, We moved the secondary mirror twice based on the results of model analyses as well as data analyses of the measured images. The whole operation took only two days. The in-orbit imaging performance thus obtained for the AKARI telescope is diffraction limited at a wavelength of 7.3 $\mu$m, slightly degraded from that expected from the laboratory measurement. The main reason for the degradaion is the additional coma aberration in the wavefront error, the cause of which is yet to be investigated. We have judged that the degradation in the imaging performance is still acceptable to achieve the major scientific goals of the AKARI mission. Additional adjustment mechanisms for the correction of two-axis tilts of a secondary mirror would significantly improve the tolerance of a telescope for such in-orbit optical misalignment, which will definitely be needed for large telescope missions in future, such as the Space Infrared Telescope for Cosmology and Astrophysics (SPICA; Kaneda et al. 2004; Onaka et al. 2005; Nakagawa and Murakami 2007).   

\bigskip

We thank all the members of the AKARI project, particularly those deeply engaged in the operation of the performance verification phase. We also express our gratitude to the IRC hardware team for their help in the operation of the telescope both on the ground and in the orbit. We are deeply grateful to R. Yamashiro and Y. Sugiyama at the Nikon Corporation for their support in the analyses of the model and the measured data. AKARI is a JAXA project with the participation of ESA.

\clearpage

\begin{figure}
\begin{center}
\FigureFile(70mm,70mm){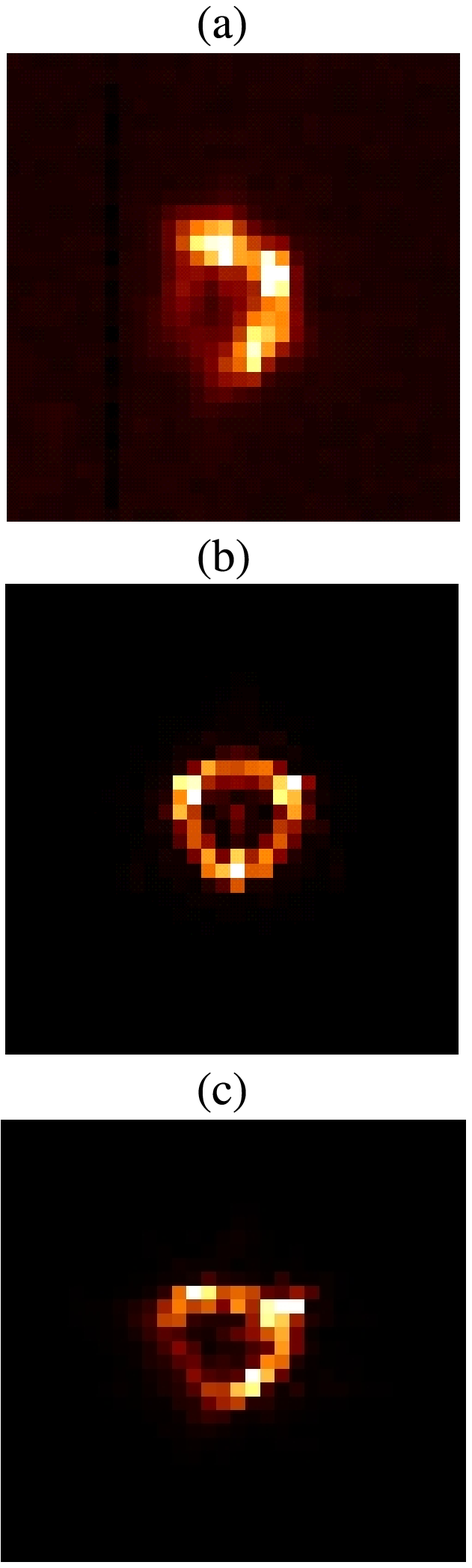}
\end{center}
\caption{(a) Star image observed in the N2 band before the focus adjustment. (b) Defocused model image of a star, and (c) that with $50''$ tilting of the primary mirror.  The image size is $32\times 32$ pixels or $47''\times 47''$.}
\end{figure}

\clearpage

\begin{figure}
\begin{center}
\FigureFile(60mm,60mm){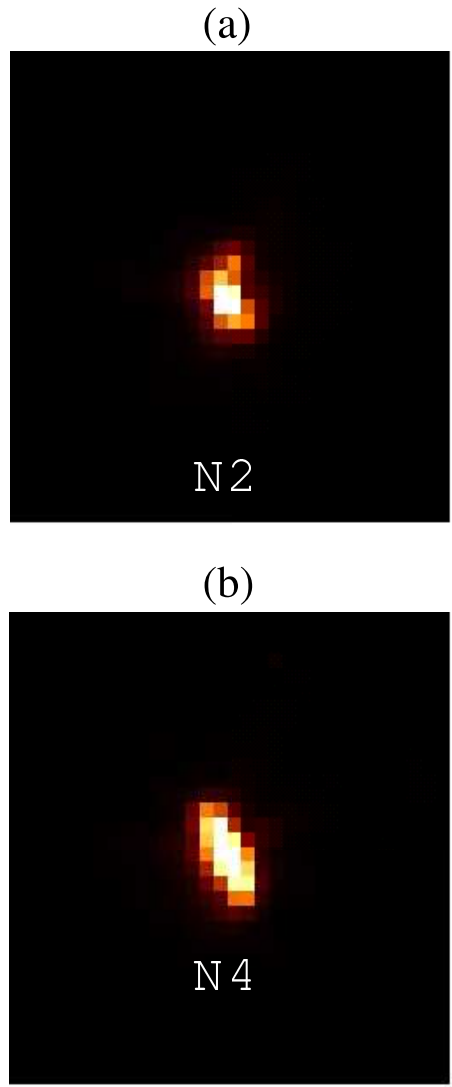}
\end{center}
\caption{Star images in (a) the N2 and (b) the N4 bands obtained after the focus adjustment of the telescope. The image size is $32\times 32$ pixels or $47''\times 47''$.}
\end{figure}

\clearpage

\begin{figure}
\begin{center}
\FigureFile(80mm,80mm){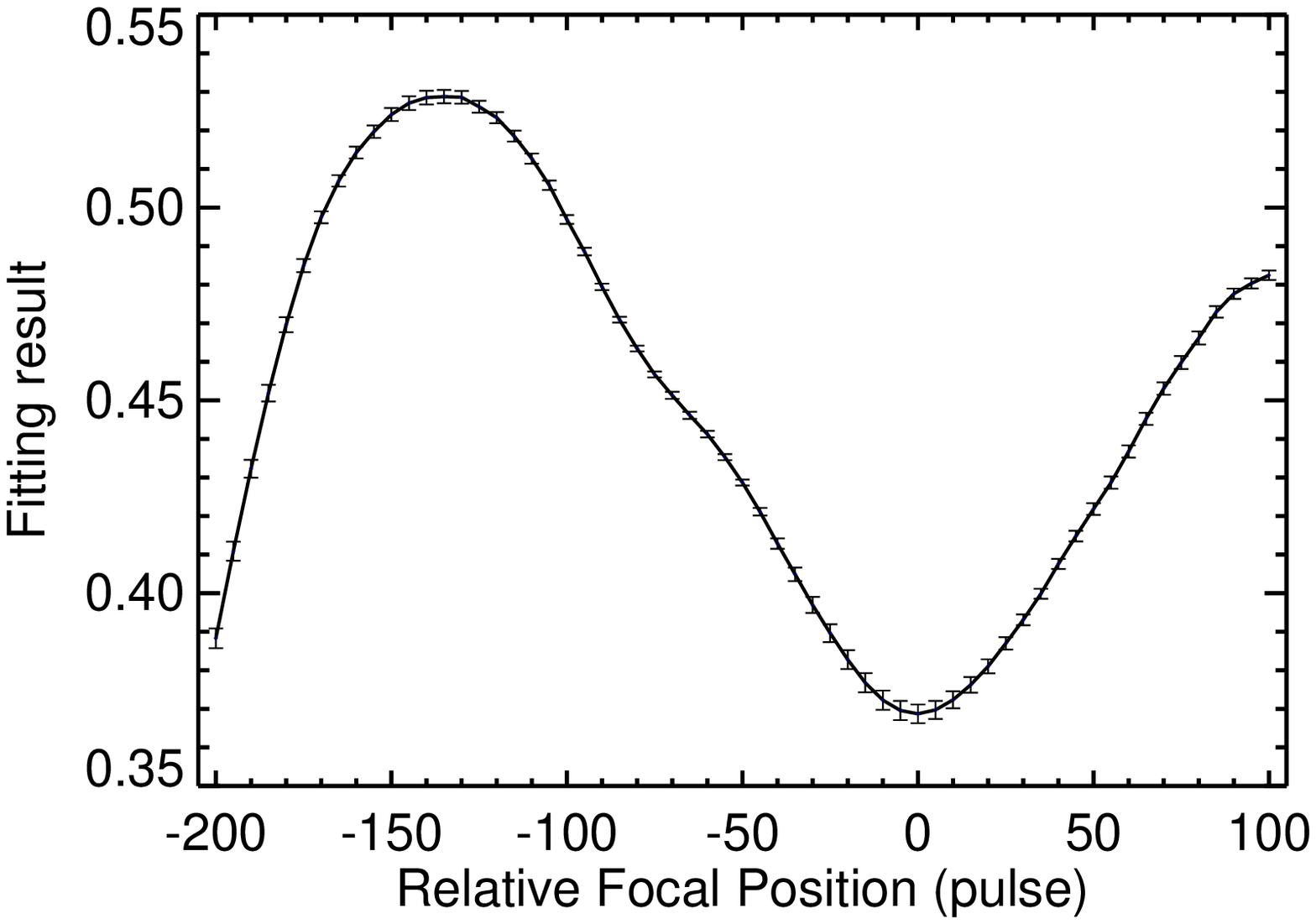}
\FigureFile(80mm,80mm){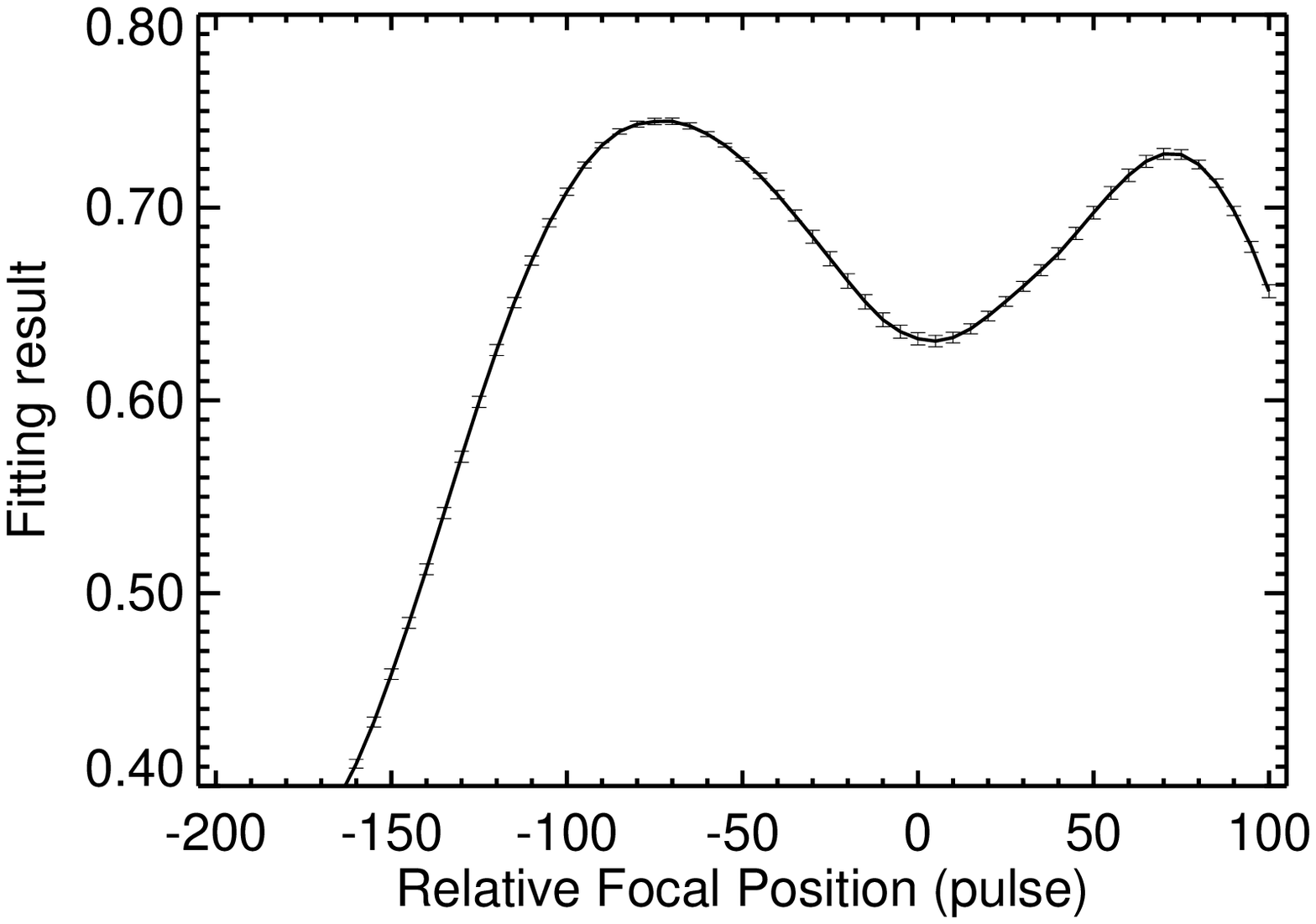}
\end{center}
\caption{Results of cross-correlation between model images and measured images (a) initially obtained in the AKARI orbit, and (b) after the first move of the secondary mirror.}
\end{figure}

\clearpage

\begin{figure}
\begin{center}
\FigureFile(80mm,80mm){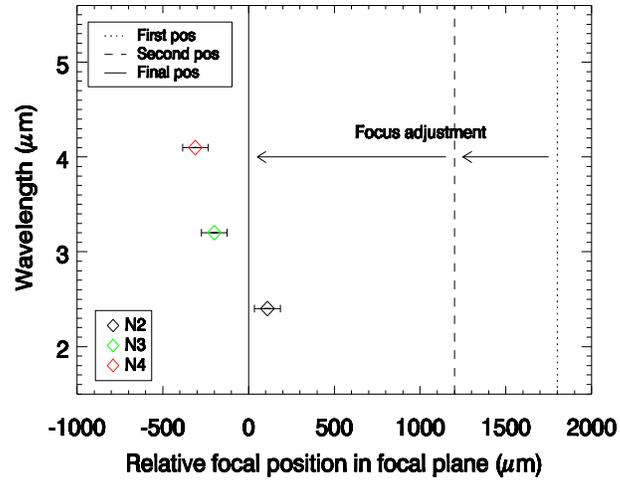}
\end{center}
\caption{Relative focal positions in the focal plane of the 3 near-IR bands.}
\end{figure}

\clearpage

\begin{figure}
\begin{center}
\FigureFile(80mm,80mm){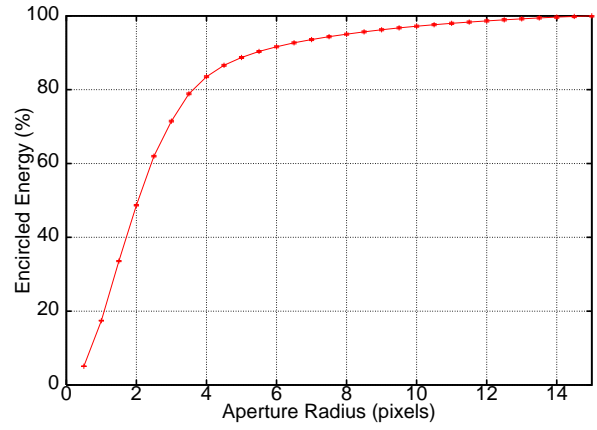}
\end{center}
\caption{Encircled energy of the PSF obtained in the N2 band.}
\end{figure}

\clearpage

\begin{table}
\caption{Characteristics of secondary focus stepping motor}
\begin{center}
\begin{tabular}{lc}
\hline\hline
Motor & PHYTRON VSS42\\
Phase current & 0.5 A\\
Torque & 0.026 Nm\\
Detent torque & 0.005 Nm\\
Step angle & 1.8 deg\\
Pulse speed & 14 pulse/sec\\
\hline
\end{tabular}
\end{center}
\end{table}

\clearpage

\begin{table}
\caption{Focus offsets between near-IR bands and sizes of star images initially obtained in the AKARI orbit}
\begin{center}
\begin{tabular}{lccc}
\hline\hline
Near-IR band  & N2 & N3 & N4\\
Focus offsets\footnotemark[$*$] & 0 & $-$0.30 $\mu$m & $-$0.42 $\mu$m \\
Image sizes\footnotemark[$\dagger$] & $3.76\pm 0.02$ & $4.16\pm 0.02$ & $4.36\pm 0.02$\\
\hline
\\
\multicolumn{4}{@{}l@{}}{\hbox to 0pt{\parbox{100mm}{\footnotesize
\par\noindent
\footnotemark[$*$] These are measured in the laboratory cryogenic test prior to the launch. The direction from the secondary to the primary mirror is taken to be positive.
\par\noindent 
\footnotemark[$\dagger$] These are defined by $r_{\rm mean}$ in equation (1), given in units of pixels.
}\hss}}
\end{tabular}
\end{center}
\end{table}

\end{document}